\documentstyle[12pt,aaspp4,epsfig]{article}
\lefthead{Aparicio et al.}
\righthead{Model CMDs for HST}

\begin{document}

\title{Model Color-Magnitude Diagrams for HST Observations of Local Group Dwarf 
Galaxies\\}

\author{A. Aparicio}
\affil{Instituto de Astrof\'\i sica de Canarias, E38200 - La Laguna, Tenerife, 
Canary Islands, Spain}

\author{C. Gallart}
\affil{Observatories of the Carnegie Institution of Washington, 813 Sta Barbara 
St., Pasadena, CA 91101, USA}

\author{C. Chiosi}
\affil{Dipartimento de Astronomia dell'Universit\`a di Padova, Vicolo 
dell'Osservatorio 5, I35122 - Padova, Italy}

\and

\author{G. Bertelli}
\affil{Consiglio delle Ricerche, Vicolo dell'Osservatorio 5, I35122 - Padova, 
Italy}

\begin{abstract}

In this paper, we discuss a method to conduct a quantitative study of the star 
formation history (SFH) of Local Group (LG) 
galaxies using Hubble Space Telescope (HST) data. This method has proven to be 
successful in the analysis of the 
SFH of the same kind of galaxies using ground-based observations. It is 
based on the comparison of observed CMDs with a set of model CMDs. The latter 
are 
computed assuming different evolutionary scenarios, and include a detailed 
simulation of observational effects.

CMDs obtained with HST are $\sim 3$ mags deeper than 
typical CMDs obtained from ground-based telescopes, allowing the observation, 
for 
all LG galaxies, of a part of the CMD that up till now had remained accessible 
only 
for 
the very nearest galaxies. A very important feature that will become accessible 
with HST is the horizontal-branch {\it plus} the red-clump. The 
distribution of stars along this structure is quite sensitive to age and 
metallicity and should provide a very important improvement in the time 
resolution of the SFH for stars older than $\simeq 2-3$ Gyr. We show and discuss 
four model CMDs which would be comparable with CMDs from deep HST observations. 
These model CMDs represent the following evolutionary scenarios corresponding to 
a 
wide range of dwarf galaxy sub-types from dI to dE: {\bf A)} a constant SFR from 
15 
Gyr ago to the present time; {\bf B)} as A), but with the SFR stopped 0.5 Gyr 
ago; 
{\bf C)} a constant SFR in the age range 10-9 Gyr and {\bf D)} as C) but in the 
age 
range 15-12 Gyr. In all four cases a range of metallicity from $Z=0.0001$ to 
$Z=0.004$ has been assumed. 

The present analysis is just a first qualitative approach to what one may expect 
to 
find in the CMDs of LG galaxies. However a complete set of model CMDs must be 
computed to analize the data for each galaxy, using the crowding effects derived 
for that particular galaxy.

\end{abstract}


\section{Introduction}

The most powerful way of studying stellar populations and the star formation 
history (SFH) of a galaxy is by means of the analysis of the distribution of 
stars 
in the 
color magnitude diagram (CMD). This kind of study can only be performed for 
systems close enough for their stars to be resolved and has usually been
limited to Local Group (LG) galaxies. It has been shown that the SFH of a LG 
galaxy can be fully determined from the analysis of its CMD, using model CMDs 
based on synthetic ones in which a careful simulation of observational effects 
is performed. Full SFHs, from the present to the very early episodes of star 
formation 
which took place more than 10 Gyr ago, have been obtained using this method for 
NGC~6822 (Gallart et al. \markcite{n682} 1996b, \markcite{n683} 1996c), Pegasus  
(Aparicio, Gallart \& Bertelli \markcite{peg2} 1996a) and LGS~3 (Aparicio, 
Gallart \& Bertelli \markcite{peg2} 1996b). 

Nevertheless, the ability to resolve the short time scale structure in the SFH necessarily worsens for older and older 
ages. Observation of the horizontal-branch (HB) 
and the red-clump (RC) of core He-burning stars is crucial to improve the time 
resolution for ages in the range from a few Gyr to the oldest ones. The 
distribution of stars in these structures will also provide constraints on the 
chemical enrichment law (CEL) at intermediate and old ages. The HB and RC have 
been 
observed for the nearest Milky Way satellites using ground-based telescopes, but 
are unreachable from the ground for more distant galaxies, including the rest of 
the LG members. However, it is expected that these structures will be observed 
in deep HST's images for any LG galaxy. These images would also contain  
important clues for young stars in the main-sequence (MS) and blue-loop 
(BL) phases, at ages up to $\sim$ 1 Gyr. As in the case of the HB and RC, these 
stars are only accessible for the Milky Way satellites from ground-based 
telescopes.

In this paper we present a guideline of how the method we have used for several 
LG 
galaxies observed from the ground, can be applied to HST observations of dwarf 
galaxies up to a distance of 1-2 Mpc. Based on model CMDs, we show the 
distribution 
of stars to be expected for four different cases of SFH: {\bf A)} a constant 
star 
formation rate (SFR) from 15 Gyr ago to the present time; {\bf B)} as A), but 
assuming the star formation ceased 0.5 Gyr ago; {\bf C)} a constant SFR at ages  
10-9 Gyr and {\bf D)} as C) but at ages 15-12 Gyr. In all four cases a range of 
metallicities from $Z=0.0001$ to $Z=0.004$ 
has been assumed. The first two cases, A) and B) may be representative of what 
can be expected for a dI galaxy or a dI-dE intermediate case such as LGS~3, 
whereas 
C) and D) might be representative of a dE galaxy in which no star formation has 
occurred since several Gyr ago.

Section~2 presents the model CMDs. In Section~3 we discuss the different 
features of those diagrams, and in Section~4 we summarize the conclusions.

\section{The Model Color Magnitude Diagrams}

The construction of a model CMD comprises two steps: {\it (i)} computation of 
a synthetic CMD and {\it (ii)} simulation of observational effects. Synthetic 
CMDs are based on a set of stellar evolutionary models covering the required 
range of masses and metallicities. The Padua stellar evolutionary library has 
been 
used here (see Bertelli et al. \markcite{padua} 1994 and references therein). 
The 
basic hypothesis under which a synthetic 
CM diagram is generated are an input SFR and CEL both as a function of time. The 
initial mass function (IMF) 
is also input information which we choose to maintain unchanged in all our 
models. 
In particular, we used the IMF derived by Kroupa, Tout \& Gilmore \markcite{imf} 
(1993). Synthetic CMDs reproduce the distribution of stars for different input 
SFR and CEL and other intervening functions and parameters which we have kept 
fixed 
here. But these diagrams cannot be compared directly with the observed CMD since 
the latter include the observational 
effects resulting mainly from stellar crowding. In spite of the much better 
spatial 
resolution of HST, crowding is expected to play the 
main role imposing a limit to the photometry, affecting also its quality, 
especially near the limiting magnitude, where the HB and the RC will be located. 
In 
this 
way, careful simulation of crowding effects is also needed to correctly 
interpret HST data, as will be shown below. Crowding effects are of three kinds: 
{\it i)} a fraction of  stars 
are lost, {\it ii)} magnitudes and color indices are systematically shifted and 
{\it iii)} affected by large external errors. The three effects are strong, non 
trivial functions of the magnitude and color index of each star and of the 
distribution of magnitudes and color indices of all the stars present in the 
galaxy (see Aparicio \& Gallart \markcite{peg1} 1995 and Gallart, Aparicio \& 
V\'\i lchez \markcite{n681} 1996a). We term synthetic CMD, the diagram prior to 
the simulation of observational effects, and model CMD the synthetic CMD after 
simulating those effects. 

For the present analysis, synthetic and model CMDs have been obtained as 
described 
in Aparicio \& Gallart 
\markcite{peg1} (1995) and Gallart et al. \markcite{n682} (1996b). Four 
synthetic CMDs have been initially computed using the ZVAR code (see Chiosi, 
Bertelli 
\& Bressan \markcite{zvar} 1988 for a discussion of an early version of it) and 
assuming a constant SFR; a linear CEL with initial metallicity 
$Z_i=0.0001$ and final metallicity $Z_f=0.004$, and the following values for 
$T_i$, $T_f$ (starting and final times for the star formation):

A) $T_i=15$ Gyr, $T_f=10^{-2}$ Gyr

B) $T_i=15$ Gyr, $T_f=0.5$ Gyr

C) $T_i=10$ Gyr, $T_f=9$ Gyr

D) $T_i=15$ Gyr, $T_f=12$ Gyr

\noindent where time is expressed in terms of age, i.e. increasing towards the 
past, 0 denoting the present moment. Using the relation by Nissen \& Shuster 
\markcite{zfe} (1991) and $Z_\odot=0.02$, the adopted range of metallicity 
$Z=0.0001$ to $Z=0.004$ corresponds to $[Fe/H]=-2.7$ to --1.1.

The observational errors have then been simulated in the synthetic CMD using the 
crowding test table obtained in the analysis of NGC~6822 by Gallart et al. 
\markcite{n681} (1996a), but with a different zero-point, such that the 
crowding effects for the hypothetical HST observations at a given stellar 
magnitude were the same as for the ground-based telescope (the 2.5 m INT at 
Roque de los Muchachos Observatory) for stars 3 magnitudes brighter. 

Figure~\ref{syn} shows the synthetic CMD for case A as a reference, since 
Figure~\ref{mod} shows the final model CMDs for cases A, B, C and D 
respectively.

\placefigure{syn}
\placefigure{mod}

NOTE FOR EDITOR: FIGURE \ref{mod} SHOULD BE FULL PAGE (TWO COLUMNS) WIDE

NOTE FOR EDITOR: PLEASE APPLY SAME REDUCTION FACTORS TO FIGURES \ref{syn} AND 
\ref{mod}

\section {Discussion}

Figure~\ref{syn} shows the synthetic CMD -prior to the simulation of crowding- 
with the 50000 stars used for case A. The main features present in the diagram 
are the 
MS, BL sequence and RGB, together with an HB 
finishing in a RC at its red edge, and an AGB red-tail structure. The single 
star at 
$I\simeq -6$, $(V-I)\simeq 2$ is a younger AGB, whereas the 
brightest stars at $(V-I)\simeq 1.0$ to 1.5 are RSGs. The tip of the RGB can be 
seen at $I\simeq -4.0$ and $(V-I) \simeq 1.2$ to 1.7. A number of red subgiant 
stars evolving from the MS to the RGB are apparent between these two sequences, 
below the HB. The synthetic diagrams for cases B, C and D lack the structures 
formed by young stars: MS, blue-loops, RSG and a part of AGBs. These diagrams 
are not shown for simplicity. 

Model CMD A (Figure~\ref{mod}a) is representative of what we should expect to 
see in a dI galaxy at about 1Mpc. A constant SFR and an early age for the 
beginning 
of star formation have been assumed as a representative SFH for a dI galaxy 
according to the results reported by Gallart et al. \markcite{n682} (1996b) for 
NGC~6822 and Aparicio et al \markcite{peg2} (1996a) for Pegasus. In that 
paper it is shown that NGC~6822 most likely began to form stars at a very early 
epoch (about 15--12 Gyr ago), from low metallicity gas, and that a SFR close to 
constant or declining in the last few Gyrs seems best to reproduce the 
observations. Figure~\ref{mod}a shows the result of simulating observational 
effects in the 
synthetic CMD of Figure~\ref{syn}: many stars in the faintest magnitude 
interval are lost and the MS and RGB are widen at their lower part. This also 
affects the HB, which becomes completely diffused into the noisy tails of the MS 
and RGB. This is an important point, meaning not only that long integration 
times will be 
needed to observe the HB and RC even for the nearest dIs, but also that the 
limiting magnitude will be imposed by crowding. This implies that if the SFH 
is to be retrieved by comparing the observed CMD with model CMDs, an accurate 
simulation of crowding effects is required as in the case of ground-based 
observations. 

The fact that the BL sequence is clearly resolved from the MS is an interesting 
feature of 
Figure~\ref{mod}a. This is not usually the case for ground-based observations, 
due to the worse crowding but also because Figure~\ref{mod}a shows less 
massive stars than common ground-based works and the gap between the MS and the 
BL 
is larger for these stars. Finally, structures 
already observed from the ground, like the red-tail, are also present 
in the CMD of Figure~\ref{mod}a and they preserve their power in the 
determination of the SFH. Note that the small field of the WFPC2 would result in 
a poor sampling of fast, hence low populated stellar evolutionary phases, like 
the upper MS and BL, the RSG and the AGB, including the red-tail itself. 
The relevance of these structures in the determination of the young SFH (MS, 
BL and RSG) and the full SFH (AGB and red-tail) is so important that 
mosaics of a number of WFPC2 frames or complementary ground-based, large field 
observations should be considered in the observational strategy if a 
quantitative derivation of the SFH is intended.

Model CMD B (Figure~\ref{mod}b) is meant to represent a limiting case of a hypothetical dI galaxy
where star formation has stopped five hundred million years ago, or a dE galaxy with a 
considerable amount of intermediate-age or even young stars. LGS~3 (Aparicio et al 
\markcite{lgs} 1996b) or Leo~I (Lee et al. \markcite{leoI} 1993) seem to be good 
examples of this intermediate type of galaxy. The very young population (MS, 
bright 
blue-loops and RSG) present in model A has disappeared here, but a large number 
of stars populate the lower part of the diagram from $(V-I)=-0.2$ to 1.2. They 
are MS and subgiant stars (but also the result of crowding which, in our 
models, is significant at this magnitude level) and they mask the HB almost 
completely so much so that it
becomes hardly distinguishable. This is an important point because the lack of a 
blue HB would be interpreted as the lack of either very old or very low 
metallicity stars. Nevertheless, if the simulation of crowding effects is 
accurate enough, a detailed comparison of the distribution of stars in the 
observed and model CMDs will show the presence of the HB and still allow 
valuable information to be obtained from it. The small group of stars over the 
red-clump, at
$(V-I)\simeq 0.85$ and $I$ from --1 to --2, is produced by stars with ages 
between 0.5 and 1 Gyr undergoing the nuclear He-burning phase.

Model CMDs C and D (Figure~\ref{mod}c and ~\ref{mod}d) are hypothetical 
representations of dE galaxies with no star formation activity over the past 9 Gyr (model C) and 12 Gyr (model D). There are several 
indications about the existence of intermediate populations in dE. For some dE 
companions 
of the Milky Way, like Carina (Mould \& Aaronson \markcite{maar} 1983; Mighell \markcite{migh} 1990 and Smecker-Hane et al. \markcite{carina} 1996), Leo~I (Lee et al. \markcite{leoI} 1993) and Leo~II (Mighell \& Rich \markcite{mighrich} 1996), the deep CMD obtained from the ground 
clearly 
show this. But for more distant systems, like the Andromeda companions, the 
controversy about the existence or not of a substantial amount of 
intermediate-age star formation is an open question (Mould \& Kristian \markcite{mk} 1990; Freedman \markcite{fre} 1992; Lee, 
Freedman \& Madore \markcite{n185} 1993; Konig et al. \markcite{konig} 1993; Armandroff et al. \markcite{arm} 1993). In this sense models~B and C can be used to test whether or not an intermediate-age population exists in a given galaxy.

Star formation occurred from 10 to 9 Gyr 
ago in model C, and from 15 to 12 Gyr ago in model D, with the same 
range of metallicities from $Z_i=0.0001$ to $Z_f=0.004$ in both cases. The RGB 
locus 
is the most apparent feature in both CMDs,
together with an HB+RC structure. In fact, the most evident difference 
between the diagrams of models C and D is the extension of the HB, solely due to 
the 
different ranges of stellar ages. Since the metallicity range can be confidently 
determined from the dispersion of the upper part of the RGB, where crowding 
effects are small (in fact there is practically no difference between the 
synthetic CMDs and the model diagrams in that area), the extension of the HB is 
crucial when determining of the age of the oldest stars in the system. Finally 
a few AGB stars are observed over the tip of 
the RGB, forming a red-tail structure in both models. 

An interesting difference between models A and B and models C and D is the width 
of the RGB, although in all cases the range of metallicities is the same. This 
is 
due to the effects of age and metallicity moving the stars' colors in opposite 
senses. Therefore, care must be taken when interpreting the width of the RGB 
as an indicator of the metallicity dispersion, unless independent proof of 
the age dispersion is available. With less deep observations than those modelled 
here, a 
hypothetical galaxy with a SFH equal to that portrayed in case B would be 
virtually 
undistinguishable from case C or D. If the absence of young stars populating the 
observed part of the CM diagram were interpreted as the galaxy being a pure 
old system, the RGB's width would be produced by the star's metallicity 
dispersion, 
leading to a much smaller value for that dispersion than the actual one, and 
therefore, an evolutionary scenario notably different from the real one would be 
inferred.

We have discussed four realistic examples of what can be expected from 
observations using HST for LG dwarf galaxies. But to quantitatively retrieve the 
SFH 
of a real galaxy would require a complete set of synthetic CMDs assuming 
different evolutionary 
scenarios (SFR + CEL basically) to be computed. Crowding effects for that 
particular set of observations should be derived and then simulated in the 
synthetic CMDs to obtain a set of model CMDs directly comparable to the observed 
CMD. However, in the light of the new photometric material that HST is producing, the four particular examples shown in this paper may help to gain 
an initial qualitative understanding of the stellar populations present in a 
galaxy.     

\section{Conclusions}

Comparison of model CMDs with observed CMDs is a powerful tool to quantitatively 
obtain the SFH of a given galaxy. Deep observations using HST will reveal two 
structures, the HB and the RC, which to date could be only observed in the 
nearby satellites 
of the Milky Way. Their analysis  will definitively allow 
the full SFH of these galaxies to be obtained accurately, with much higher 
resolution towards old ages than the 
SFH obtained up till now from ground-based observations. 

To illustrate what can 
be obtained from long integrations with HST for galaxies at a distance of about 
1 Mpc, four examples of model CMDs, for four different SFHs, have been presented 
and 
discussed. These model CMDs can be considered as first approachs to the CMD of a 
dI galaxy (model A), an intermediate case between dI and dE (model B) and two 
cases for dE galaxies with no star formation since several Gyrs ago (models C 
and D). The HB+RC structure is 
clearly visible in cases C and D and promises to be particularly powerful in 
the detailed determination of the intermediate-age and old SFH. Long 
integration times seem necessary to detect them with good 
signal to noise ratio. However the HB is hardly visible in models A and B where 
they appear embedded in the diffuse cloud produced by crowded MS, RGB and 
subgiant stars. Nevertheless, if the simulation of crowding effects is 
accurate enough, a detailed comparison of the distribution of stars in the 
observed and model CMDs may still yield valuable information.

\newpage
\figcaption[syn.eps]{Synthetic CMD for constant SFR, $T_i=15$ Gyr, $T_f=10^{-2}$ 
Gyr, $Z_i=0.0001$ and $Z_f=0.004$.
\label{syn}}

\figcaption[mod.eps]{CMDs for the four model SFH discussed. Constant SFR and 
initial and final metallicities $Z_i=0.0001$, $Z_f=0.004$ are common to the four 
models. The time marking the beginning of the star formation is $T_i=15$ Gyr for 
models 
A, B and D, and $T_i=10$ Gyr for model C. The time marking the end of star 
formation is 
$T_f=10^{-2}$ Gyr for model A; $T_f=0.5$ Gyr for model B; $T_f=9$ Gyr for model 
C and $T_f=12$ Gyr for model D. $T=0$ is the present time.
\label{mod}}

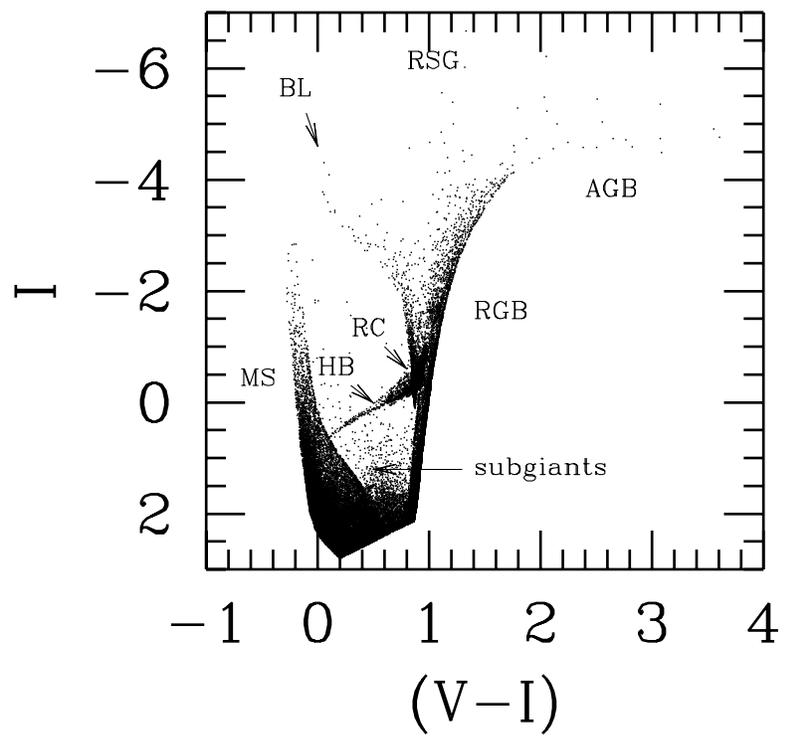
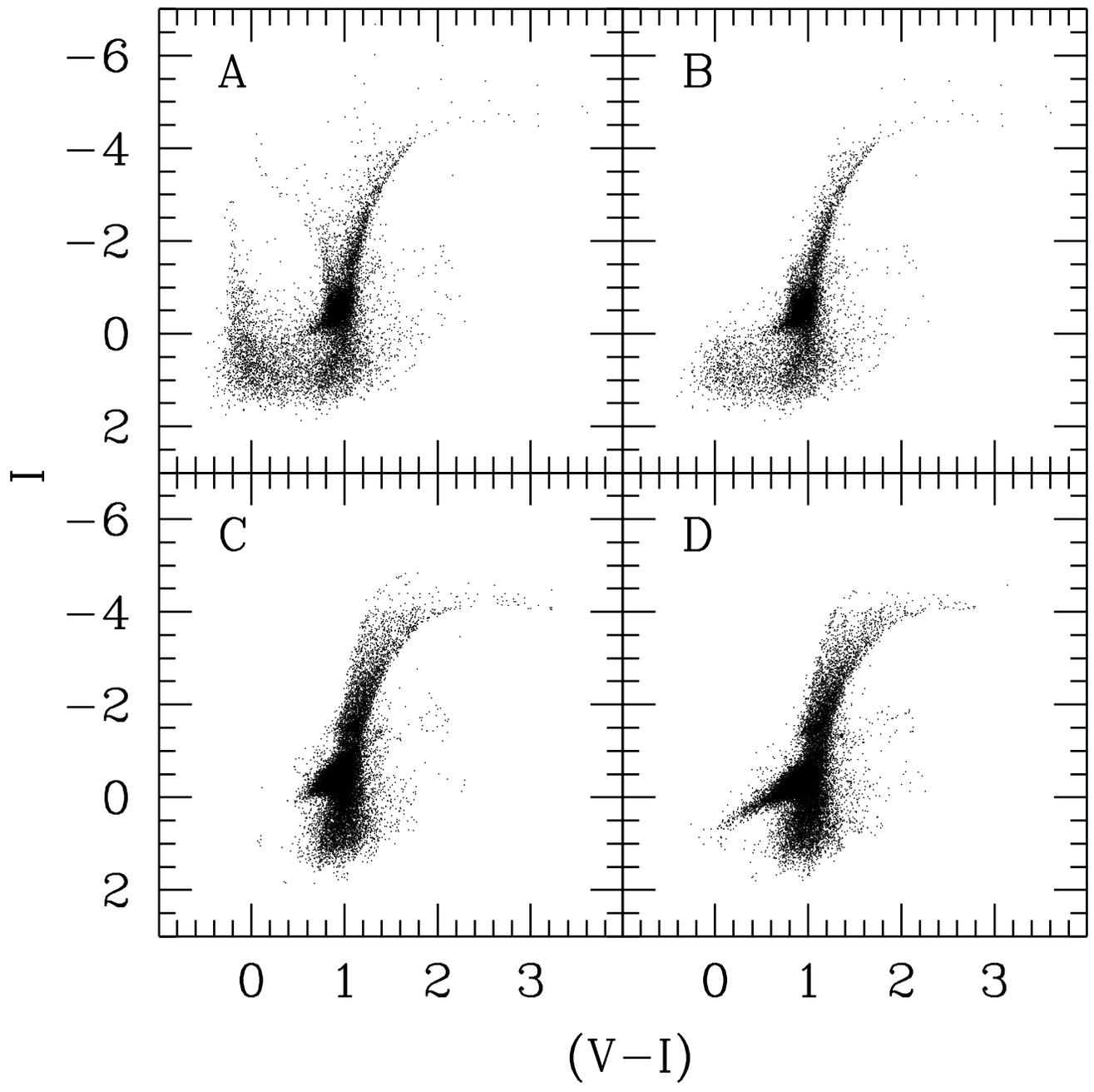
\end{document}